\documentclass[apjl]{emulateapj}
\slugcomment{(To appear in the ApJ Letters)}
\newcommand{\Herschel}{{\it Herschel}}
\newcommand{\CI}{[C\,{\sc i}]}
\newcommand{\NII}{[N\,{\sc ii}]}

\begin{document}

\shorttitle{Warm Molecular Gas in LIRGs}
\shortauthors{Lu et al.}

\title{Warm Molecular Gas in Luminous Infrared Galaxies\footnotemark[1]}
\footnotetext[1]{Based on \Herschel\ observations. \Herschel\ is an ESA space observatory with science 
instruments provided by European-led Principal Investigator consortia 
and with important participation from NASA.} 

\author{N. Lu\altaffilmark{2}, Y. Zhao\altaffilmark{2, 3, 4}, C. K. Xu\altaffilmark{2}, Y. Gao\altaffilmark{3, 4}, 
L. Armus\altaffilmark{5}, J. M. Mazzarella\altaffilmark{2}, K. G. Isaak\altaffilmark{6}, A.~O.~Petric\altaffilmark{7, 8},
V. Charmandaris\altaffilmark{9,10,11}, T.~D\'iaz-Santos\altaffilmark{5},  A. S. Evans\altaffilmark{12, 13}, 
J. Howell\altaffilmark{2},  P. Appleton\altaffilmark{2}, H. Inami\altaffilmark{14},  K. Iwasawa\altaffilmark{15}, 
J. Leech\altaffilmark{16}, S. Lord\altaffilmark{2},  D.~ B.~ Sanders\altaffilmark{17}, B. Schulz\altaffilmark{2}, 
J. Surace\altaffilmark{5}, P. P. van der Werf\altaffilmark{18}}
\altaffiltext{2}{Infrared Processing and Analysis Center, California Institute of Technology, MS 100-22, Pasadena, CA 91125, USA; lu@ipac.caltech.edu}
\altaffiltext{3}{Purple Mountain Observatory, Chinese Academy of Sciences, Nanjing 210008, China}
\altaffiltext{4}{Key Laboratory of Radio Astronomy, Chinese Academy of Sciences, Nanjing 210008, China}
\altaffiltext{5}{Spitzer Science Center, California Institute of Technology, MS 220-6, Pasadena, CA 91125, USA}
\altaffiltext{6}{ESA Astrophysics Missions Division, ESTEC, PO Box 299, 2200 AG Noordwijk, The Netherlands}
\altaffiltext{7}{Gemini Observatory, 670 N. A$^{\prime}$ohoku Place, Hilo, HI 96720, USA}
\altaffiltext{8}{Astronomy Department, California Institute of Technology, Pasadena, CA 91125, USA}
\altaffiltext{9}{Department of Physics, University of Crete, GR-71003 Heraklion, Greece}
\altaffiltext{10}{IAASARS, National Observatory of Athens, GR-15236, Penteli, Greece}
\altaffiltext{11}{Chercheur Associ\'e, Observatoire de Paris, F-75014, Paris, France}
\altaffiltext{12}{Department of Astronomy, University of Virginia, 530 McCormick Road, Charlottesville, VA 22904, USA}
\altaffiltext{13}{National Radio Astronomy Observatory, 520 Edgemont Road, Charlottesville, VA 22903, USA}
\altaffiltext{14}{National Optical Astronomy Observatory, Tucson, AZ 85719, USA}
\altaffiltext{15}{ICREA and Institut de Ci\`encies del Cosmos (ICC), Universitat de Barcelona (IEEC-UB), Mart\'i i Franqu\`es 1, 08028, Barcelona, Spain}
\altaffiltext{16}{Department of Physics, University of Oxford, Denys Wilkinson Building, Keble Road, Oxford, OX1 3RH, UK}
\altaffiltext{17}{University of Hawaii, Institute for Astronomy, 2680 Woodlawn Drive, Honolulu, HI 96822, USA}
\altaffiltext{18}{Leiden Observatory, Leiden University, PO Box 9513, 2300 RA Leiden, The Netherlands}


\begin{abstract}
We present our initial results on the CO rotational spectral line energy 
distribution (SLED) of the $J$ to $J$$-$1 transitions from $J=4$ up to $13$
from \Herschel\ SPIRE spectroscopic observations of 65 luminous 
infrared galaxies (LIRGs) in the Great Observatories All-Sky LIRG Survey 
(GOALS).  The observed SLEDs change on average from one peaking at $J \le 
4$ to a broad distribution peaking around $J \sim\,$6$-$7 as the {\it IRAS}
60-to-100\,\micron\ color, $C(60/100)$, increases.  
However, the ratios of a CO line luminosity to the total infrared luminosity,
$L_{\rm IR}$, show the smallest variation for $J$ around 6 or 7.  This suggests
that, for most LIRGs, ongoing star formation (SF) is also responsible for 
a warm gas component that emits CO lines primarily in the mid-$J$ regime 
($5 \lesssim J \lesssim 10$).   As a result, the logarithmic ratios of 
the CO line luminosity summed over CO\,(5$-$4), (6$-$5), (7$-$6), (8$-$7) and 
(10$-$9) transitions to $L_{\rm IR}$,  $\log R_{\rm midCO}$,  remain largely 
independent of $C(60/100)$, and show a mean value of $-4.13$ ($\equiv 
\log R^{\rm SF}_{\rm midCO}$) and a sample standard deviation of only 0.10 
for the SF-dominated galaxies.  Including additional galaxies from the literature,
we show, albeit with small number of cases, the possibility that 
galaxies, which bear powerful interstellar shocks unrelated to the current SF, 
and galaxies, in which an energetic active galactic nucleus contributes 
significantly to the bolometric luminosity, have their $R_{\rm midCO}$ 
higher and lower than $R^{\rm SF}_{\rm midCO}$, respectively.
\end{abstract}


\section{INTRODUCTION} \label{sec1}

Luminous Infra-Red Galaxies [LIRGs, defined as $L_{\rm IR}$(8$-$1000\,\micron)
$\ge 10^{11}\,$L$_{\odot}$] dominate the cosmic star formation (SF) at $z > 1$
(Le Fl\'och et al.~2005; Magnelli et al.~2009).  The local counterparts are 
all known to be rich in molecular gas (Sanders \& Mirabel 1996).
While the CO\,(1$-$0) line has been widely used to trace total 
molecular gas content, SF occurs in the denser parts of molecular gas as 
evidenced by correlations between $L_{\rm IR}$ and dense gas tracers, 
such as HCN\,(1$-$0) (e.g., Gao \& Solomom~2004; Wu et al.~2005).  SF is 
expected to heat up the molecular gas substantially.  The resulting warm 
gas can be better traced by mid-$J$ CO line transitions such as CO\,(6$-$5),
which has a critical density of $\sim$3$\times 10^5\,$cm$^{-3}$ and an 
excitation temperature of $\sim$116\,K (Carilli \& Walter 2013), hinted 
already by limited ground-based CO data (e.g., Bayet et al.~2009). 

We have carried out a survey of warm molecular gas for a 
sample of 125 LIRGs belonging to the Great Observatories All-Sky LIRG
Survey (GOALS; Armus et al.~2009), by obtaining line fluxes of the CO 
$J$ to $J$$-$1 transitions from $J=4$ up to 13 with the SPIRE Fourier 
Transform Spectrometer (FTS; Griffin et al.~2010) on-board \Herschel\ 
(Pilbratt et al.~2010).  The other targeted lines include \NII\ 205
\micron, \CI\ 609 \micron\ and \CI\ 370 \micron.
The program and observational data on individual galaxies will be given
in full in a future paper (Lu et al 2014; in preparation).  An earlier 
paper (Zhao et al.~2013) presented the initial results on how 
the \NII\ 205 \micron\ line correlates with the SF rate.  In this
paper we report our initial findings on the observed CO line emission
for 65 sample galaxies from the {\it early} \Herschel\ observations. 
We provide a brief description of our survey program and data reduction 
in \S2, and present our results and discuss physical implications on 
gas and dust heating in \S3.

\section{The Sample, Observations and Data Reduction} \label{sec2}

Our full FTS sample is a flux-limited subset of the GOALS sample, which 
contains 202 LIRGs down to $f_{\nu}(60\micron) = 5.24\,$Jy, by further 
satisfying $F_{\rm IR}$(8$-$1000\,\micron) $> 6.5\times 10^{-13}\,$W\,m$^{-2}$, 
where $F_{\rm IR}$ is as defined in Sanders \& Mirabel (1996) and 
the conversion between $F_{\rm IR}$ and $L_{\rm IR}$ used the luminosity 
distance given in Table~1 in Armus et al.~(2009).  Galaxies in a pair 
were treated separately with their individual $L_{\rm IR}$ derived as 
in D\'iaz-Santos et al.~(2010).  Our FTS sample contains 7 galaxies
with $L_{\rm IR} > 10^{12} L_{\odot}$.   Ninety-three galaxies were 
eventually observed by us (program ID: OT1$_-$nlu$_-$1) using the high 
resolution, sparse mode centered on each galaxy.  
The remainder have archived FTS data.  Our own program emphasized detections 
of the mid-$J$ CO lines (i.e., $5 \lesssim J \lesssim 10$).  The on-target 
integration times range from 1,332 to 7,992~sec, set to detect 
the anticipated CO\,(6$-$5) flux at S/N $> 5$.

The 65 galaxies chosen for this paper are practically point-like sources 
with respect to the SPIRE beams (see Zhao et al.~2013), including 38 from
own observations, 24 archived observations originally from the HerCULES program 
(PI: P. van der Werf; van der Werf et al.~2010),  UGC\,05101 and NGC\,7130
from Pereira-Santaella et al (2013), and Arp\,220 from Rangwala et al.~(2011).
The SPIRE data were reduced homogeneously using 
{\Herschel} Interactive Processing Environment (HIPE), version 9 (Ott 2010), 
which offers a line flux accuracy better than 10\%.  
The final spectrum was extracted from the two central detectors (SSWD4 and 
SLWC3).  As an example,  Fig.~1 shows 
the spectrum of {\it IRAS}\,17578-0400, a starburst with $\log L_{\rm IR}/L_{\sun}
= 11.40$ and {\it IRAS} 60-to-100\,\micron\ flux density ratio (also referred 
to as FIR color), $C(60/100) \sim 0.83$, with the fitted CO, \CI\ ad \NII\ 
lines marked.

\begin{figure}[t]
\centering
\includegraphics[width=.50\textwidth, bb=15 0 598 476]{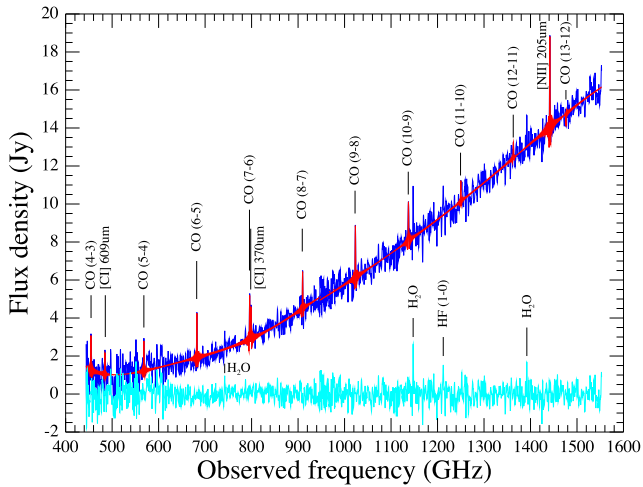}
\caption{
The observed SPIRE spectrum (in blue) of {\it IRAS}\,17578-0400 
(RA $= 18^{\rm h}00^{\rm m}31.86^{\rm s}$, Dec $= -4$\arcdeg00\arcmin53.3$''$; J2000), a starburst 
with $\log L_{\rm IR}/L_{\sun} = 11.40$.
The composite fit of the continuum and the selected lines in CO, \CI\ and \NII\ 
(see the text) is shown in red.  Other detected molecular lines 
were not fit here, but marked in the residual spectrum (in cyan).
}
\label{Fig1}
\end{figure}

\begin{figure}
\centering
\includegraphics[width=.62\textwidth,, bb=20 144 600 718]{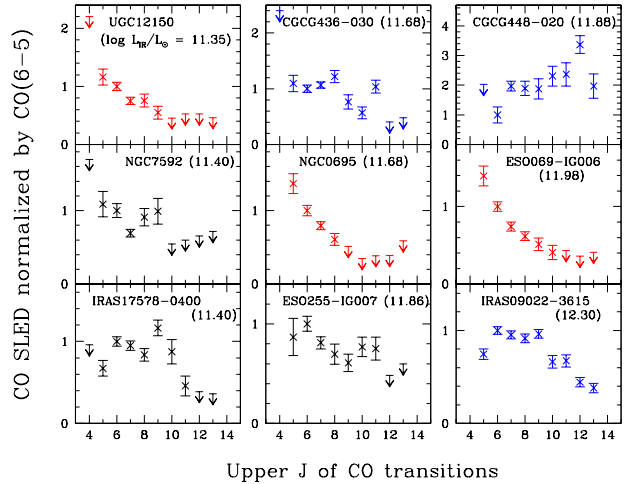}
\vspace{-1.65in}
\caption{
Examples of the observed CO SLEDs, arranged in order of increasing 
$\log L_{\rm IR}/L_{\sun}$ (given in parentheses).   
The undetected CO lines are shown with their 3$\,\sigma$ upper limits. 
Each CO SLED is normalized by the CO\,(6$-$5) flux. The SLEDs are color
coded according to the FIR color, $C(60/100)$ (see the text): red for 
$0.5 < C(60/100) < 0.6$, black for $0.75 < C(60/100) < 0.85$, 
and blue for $1.0 < C(60/100) < 1.2$.  Note that some galaxies have 
no CO\,(4$-3$) data due to redshift. 
}
\label{Fig2}
\end{figure}

We used the spectrum fitter in 
HIPE to simultaneously fit a polynomial (of order 5) to the continuum 
and SINC functions in frequency to the targeted lines.  
For each SINC profile, only its FWHM was fixed at 1.44\,GHz.  A line 
was regarded as being detected if (a) the fitted line flux is greater
than 3 times the r.m.s.~noise, measured within a frequency interval 
of 20\,FWHM wide, centered on the line frequency in the residual 
spectrum, and (b) its inferred line velocity 
difference with the (usually) brightest \NII\ 205 \micron\ line is within 
the expected uncertainty based on the spectral resolution and data 
sampling frequency.  The SINC line profiles were adequate for all the CO 
lines detected in nearly all sources. The only clear exception was
NGC\,6240, for which the broad CO lines are partially resolved and were 
fit with a Gaussian convolved profile using the algorithm described in 
Zhao et al.~(2013).

\section{Results and Discussion} \label{sec3}

\subsection{CO Spectral Line Energy Distribution} \label{sec3.1}

Fig.~2 shows 9 examples of the observed CO spectral line energy distributions
(SLEDs) from our own observations, in order of increasing $L_{\rm IR}$.
These plots illustrate that $L_{\rm IR}$ is not the best predictor for 
the SLED shapes.  For example, CGCG\,448-020 and ESO\,069-IG006 are similar in 
$L_{\rm IR}$,  but have very different CO SLEDs: The SLED of the former peaks at 
$J > 10$ while that of the latter peaks at $J \lesssim 5$.   This difference
is mainly due to their different dust temperatures:  $C(60/100) = 1.08$
for CGCG\,448-020 {\it vs.} $0.56$ for ESO\,069-IG006.  To show how $C(60/100)$, 
which measures the average intensity of the dust heating radiation field 
(Tuffs \& Popescu 2003), correlates with the SLED shape, we color-coded 
the SLEDs in Fig.~2 on $C(60/100)$.  Almost all the red-colored SLEDs, with $C(60/100) 
\le 0.6$, peak at $J \lesssim 5$ with the line fluxes decreasing rapidly as $J$ 
increases. In contrast, the blue-colored SLEDs, with $C(60/100) > 1.0$, 
are dominated by a broad distribution over $5 \lesssim J \lesssim 10$.  This 
suggests that it is the intensity of the radiation field, rather than 
$L_{\rm IR}$, which determines the SLED shape.

Fig.~3 plots the logarithmic CO line luminosity, normalized by $L_{\rm IR}$, 
as a function of $C(60/100)$ for CO\,(4$-$3), CO\,(6$-$5), CO\,(7$-$6) and 
CO\,(12$-$11).  An undetected CO line is shown with its 
3$\,\sigma$ upper limit.
The average fractional AGN contribution to the bolometric luminosity, 
$f_{\rm AGN}$, has been derived for most of the GOALS galaxies.  This 
was from a set of the mid-IR diagnostics based on [Ne\,{\sc v}]/[Ne\,{\sc ii}], 
[O\,{\sc iv}]/[Ne\,{\sc ii}], continuum slope, PAH equivalent width 
and the diagram of Laurent et al.~(2000), following the prescriptions in 
Armus et al.~(2007) (see also Petric et al.~2011; Stierwalt et al.~2013).  
Of the galaxies used here, only two have $f_{\rm AGN} > 40\%$ (i.e., $\sim$56\% 
for Mrk\,231 and $\sim$60\% for IRASF\,05189-2524). The remainder all have 
$f_{\rm AGN} < 40\%$, including only 3 galaxies with $f_{\rm AGN}$ between 30 
and 40\%, and are thus SF dominated.  In Fig.~3, the two galaxies with 
$f_{\rm AGN} > 40\%$ are further enclosed by a circle.

All the plots in Fig.~3 span 1.6 dex vertically so that the sample dispersions
can be visually compared.  The dotted line in each plot helps identifying 
the most energetic CO line at any given $C(60/100)$ by noticing that a galaxy
lies on a vertical line across the plots. 
Fig.~3 shows that as $C(60/100)$ increases, the overall CO gas
gets warmer:  The slope of the linear fit (i.e., the dashed line in each plot) to 
the detections {\it only} (but excluding the outlier NGC\,6240) increases 
monotonically from about -0.8 for CO\,(4$-$3) to +0.7 for CO\,(12$-$11).  
Both plots of CO\,(4$-$3) and CO\,(12$-$11) contain quite a few upper limits.
We performed two tests on whether the fitted slope could be substantially
biased due to the non-detections: (a) a full regression including the upper 
limits based on a survival analysis outlined by Isobe et al.~(1986), and 
(b) using only the detections from the 27 galaxies with the more sensitive 
CO data from the \Herschel\ archive, for which there are fewer upper limits. 
The results from both tests confirm, within the fit uncertainties, that 
the slopes derived from the fits to the detections are representative of 
the entire sample. 

A more significant message from Fig.~3 is that the ratio of CO\,(6$-$5) or
CO\,(7$-$6) to the IR emission is rather constant over the range of 
$C(60/100)$ probed. This argues for at least two gas components:  A warm 
component, which emits the CO lines primarily in mid-$J$ (i.e., $5 
\lesssim J \lesssim 10$) and correlates best with the dust emission, 
is mainly responsible for the observed constant ratio seen around $J = 6$
or 7.  Since the dominant heating source for 
the IR emission is current SF, the same ongoing SF should be also responsible 
for this warm CO gas component.  The other component is a ``cold'' 
and predominantly less dense gas that emits CO lines primarily at 
$J \lesssim 4$ and is {\it not directly} related to current SF.  As 
$C(60/100)$ increases, the warm gas component becomes more energetic, 
resulting in the trend seen in Figs.~2 and 3.   As we argue in \S3.2, 
in galaxies with a prominent AGN, there could also be a meaningful 
third component of hot and likely denser gas, which emits CO lines primarily 
at $J \gtrsim 10$ and is thus not sampled adequately by the SPIRE 
spectrometer.

\begin{figure}[t]
\centering
\includegraphics[width=.54\textwidth]{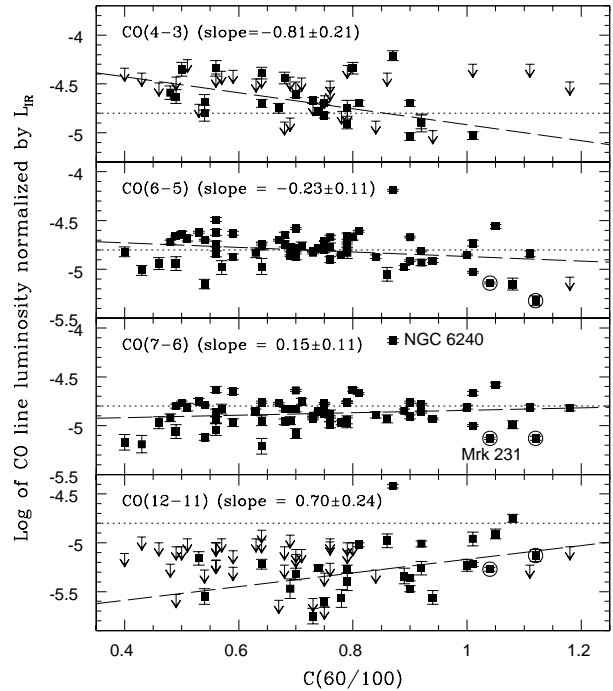}
\caption{
Plots of the logarithmic luminosities of selected CO lines, each normalized 
by $L_{\rm IR}$, as a function the FIR color for our sample.  The CO 
transition is labelled in each plot.  The two galaxies with energetic AGNs 
(see the text) are enclosed by circles.  The arrows indicate the 3$\,\sigma$ 
upper limits of an undetected line.  In each plot, the dashed line is 
the least-squares fit to the detections only (but excluding NGC\,6240), of 
which the slope is given.  The dotted line in each plot is fixed at $-4.8$.
Both NGC\,6240 and Mrk\,231 are marked. Note that some galaxies lack 
the CO\,(4$-$3) data due to redshift.
}
\label{Fig3}
\end{figure}

\begin{figure}[t]
\centering
\includegraphics[width=.55\textwidth, bb = 20 144 600 718]{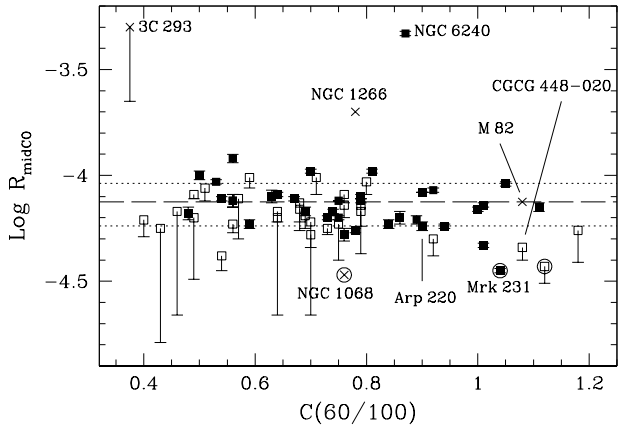}
\vspace{-1.5in}
\caption{
Plot of $\log$ of $R_{\rm midCO}$, the ratio of the combined luminosity from 
the representative mid-$J$ CO lines of CO\,(5$-$4), (6$-$5), (7$-$6), (8$-$7)
and (10$-$9) to $L_{\rm IR}$, as a function of the FIR color for our sample 
galaxies (solid or open squares) and 4 additional galaxies from the literature
(crosses).  The galaxies with a significant AGN are further enclosed by 
a circle.  For those with one or two CO lines below the 3$\,\sigma$ detection 
threshold,  they are shown by an open square or cross that represents the ratio
when the 3$\,\sigma$ upper limits are used for the undetected lines, with 
a one-sided line extending to the ratio as though the undetected lines had 
zero fluxes.  The sample average and standard deviation for the SF-dominated 
galaxies in our own sample (excluding NGC\,6240) are 
marked by the dashed and dotted lines, respectively (see the text). A few 
galaxies are labelled here and discussed in the text.
}
\label{Fig4}
\end{figure}

\vspace{0.2in}
\subsection{Molecular Gas and Dust Heating} \label{sec3.2}

Various heating mechanisms have been considered in the literature for warm 
CO line emission in galaxies, including far-UV photon heating from massive
stars (hereafter referred to as photon dominated region or PDR scenario) 
(e.g., IC\,342, Rigopoulou et al.~2013), 
X-ray photon heating from AGNs (referred to as X-ray dominated region or XDR
scenario) (e.g., Mrk\,231, 
van der Werf et al. 2010; NGC\,1068, Spinoglio et al.~2012), heating by 
cosmic-rays enhanced from supernovae (e.g., NGC\,253; Bradford et al.~2003),
and interstellar shocks (e.g., Flower \& Pineau Des For\^ets 2010). 
The shock scenario has been favored in many cases, including M\,82 
(Kamenetzky et al.~2012), Arp\,220 (Rangwala et al.~2011), NGC\,891 
(Nikola et al.~2011), NGC~253 (Rosenberg et al.~2014),  NGC\,6240 
(Meijerink et al.~2013), and NGC\,1266 (Pellegrini et al.~2013).
In this paper, we further divide the shock scenario into two categories:  
(A) shocks associated with processes derived from current SF, such as 
supernovae and stellar winds;  and (B) shocks that derive from energy 
sources other than current SF, such as those associated with AGN-driven
gas outflows, radio jets, or galaxy-galaxy collision.  While studies 
of XDR point to CO SLEDs peaking well beyond mid-$J$ (e.g., Spaans \& 
Meijerink~2008), all the other heating mechanisms can, in principle,
produce similar CO SLEDs within the mid-$J$ regime, making them difficult
to be differentiated based on SPIRE CO data alone.

In Fig.~4, we sum over the mid-$J$ CO lines of $J =$ 5, 6, 7, 8 and 10,
and plot in log the resulting luminosity to $L_{\rm IR}$ ratio, $R_{\rm midCO}$, 
as a function of $C(60/100)$. [The CO\,(9$-$8) line was left out because 
the FTS spectra are usually noisier around its frequency.]
Our sample galaxies are in solid or open squares, with the latter
representing those with one or more of the CO lines undetected. 
For these galaxies, we present the possible range for $R_{\rm midCO}$
between the 0 and 3$\,\sigma$ values for the undetected lines.  
In most of these cases, the actual ratio should be much closer to the 
upper end of the range as most of the undetected lines in the mid-$J$ 
regime have a fitted line flux not far below the 3$\,\sigma$ threshold.  
Overall, $R_{\rm midCO}$ shows little systematic dependence on 
$C(60/100)$, except for (i) $C(60/100) \gtrsim 1$, where more data 
points (e.g., Mrk\,231) tend to lie below the sample mean, and (ii) 
the clear outlier NGC\,6240. As comparison cases, we also plot in Fig.~4 
four additional galaxies indicated by crosses.  
Three are from the references listed above:  M\,82, an archetypical 
starburst; NGC\,1266, a disk galaxy with possible AGN-driven molecular
outflows (Alatalo et al.~2011); and NGC\,1068, a well-known Seyfert
galaxy with its CO line emission detected up to $J = 30$ (Hailey-Dunsheath
et al.~2012) and $f_{\rm AGN} \sim 50\%$ (Telesco \& Decher~1988).  
M\,82 is slightly extended and its $L_{\rm IR}$ used here was limited to 
within the same beam size as for the CO SLED.   The fourth object 
is 3C\,293, a radio galaxy rich in warm H$_2$, possibly excited by 
radio jet-driven shocks (Ogle et al.~2010).   From the archival SPIRE 
spectrum of 3C\,293 (Obs.~ID: 1342238242; PI: P. Papadopoulos), we derived the fluxes 
of 1.69, 2.91 and 2.14$\times 10^{-18}$ W\,m$^{-2}$ for CO\,(6$-$5), 
CO\,(7$-$6) and CO\,(8$-$7), respectively, all at S/N $\gtrsim 3$, 
but only the 3$\,\sigma$ upper limits of 5.49 and 3.15$\times 
10^{-18}$ W\,m$^{-2}$ for CO\,(5$-$4) and CO\,(10$-$9), respectively. 
Its $F_{\rm IR} \approx 3.07\times 10^{-14}$ W\,m$^{-2}$, based on 
an 12\,\micron\ flux from Siebenmorgen et al.~(2004) and {\it IRAS} 
flux densities at 25, 60 and 100\,\micron\ from Golombek, Miley 
\& Beugebauer~(1988). The galaxy 3C\,293 is shown in Fig.~4 by 
its possible range in $R_{\rm midCO}$.  In both NGC\,1266 and 3C\,293, 
our estimated $f_{\rm AGN} < 10\%$, based on [O\,{\sc iv}]/[Ne\,{\sc ii}] 
from Dudik, Satyapal \& Marcu (2009) or Ogle et al.~(2010).

The relatively small sample scatter in Fig.~4 for the SF-dominated LIRGs 
(except for NGC\,6240) suggests convincingly that the current SF is also 
the main heating source for the mid-$J$ CO line emission.  If we exclude
NGC\,6240 on one side of the main locus and the two AGNs from our sample
on the other side, the average and sample standard deviation of 
$\log R_{\rm midCO}$ from the remaining detections in our {\it own sample}
are $-4.13$ ($\equiv \log R^{\rm SF}_{\rm midCO}$; i.e., the dashed line 
in Fig.~4) and $0.10$ ($\equiv \log \sigma_{\rm SF}$; marked by the two 
dotted lines in Fig.~4), respectively.

NGC\,6240 has $R_{\rm midCO}/R^{\rm SF}_{\rm midCO} \sim 6.3$, which 
is $\sim$8\,$\log \sigma_{\rm SF}$ above $\log R^{\rm SF}_{\rm midCO}$.
The starburst superwinds in NGC\,6240 are believed to power 
the large-scale diffuse ionized gas (e.g., Heckman et al.~1987) and 
shock-excited H$_{2}$ line emission (e.g., Max et al.~2005).  
It is not clear how a category-A shock case can lead to such
a high $R_{\rm midCO}$.  The energy output of stellar winds should 
scale in certain way with that of the far-UV photons powering 
$L_{\rm IR}$, resulting in $R_{\rm midCO}$ around some characteristic 
value.  Indeed, all the other superwind galaxies (e.g., Arp\,220 
and M\,82) show a ``normal'' $R_{\rm midCO}$.  The fact that no galaxies 
are seen at intermediate $R_{\rm midCO}$ values argues against 
the hypothesis that NGC\,6240 represents an optimal wind-gas coupling
efficiency.  Therefore, we can not rule out the possibility that some 
forms of category-B shocks might also be at work in the nuclear region
of NGC\,6240, where the gas is highly turbulent (Tacconi et al.~1999) 
and the heating source of the warm H$_2$ emission is still controversial,
with arguments for (e.g., Ohyama et al.~2000) and against (e.g., 
Feruglio et al.~2013) a superwind case.  A category-B shock scenario 
can in principle explain a higher $R_{\rm midCO}$ as there is little far-UV 
counterpart.  Indeed, both NGC\,1266 and 3C\,293, where the warm CO 
emission might be enhanced respectively by the AGN- and radio jet-driven
shocks, show a higher $R_{\rm midCO}$ in Fig.~4.

On the other hand, both Mrk\,231 and NGC\,1068 have a lower $R_{\rm midCO}$
in Fig.~4.   They both have significant hot CO gas emitting at $J > 10$ 
(Gonz\'alez-Alfonso et al.~2014; Hailey-Dunsheath et al.~2012), likely 
associated with XDR (van der Werf et al.~2010;  Spinoglio et al.~2012). 
Furthermore, the model prediction that the XDR-associated CO emission 
occurs mainly at higher $J$ levels is also supported by the observation 
that, with comparable $C(60/100)$, M\,82 and Mrk\,231 have almost identical 
CO SLEDs for $J \lesssim 10$ (Pereira-Santaella et al.~2013).
These results suggest that an $R_{\rm midCO}$ significantly lower than 
$R^{\rm SF}_{\rm midCO}$ may be a result of missing most of the XDR-associated
CO line cooling at $J > 10$.    However, not all 
the galaxies with a lower CO/IR ratio in Fig.~4 have a powerful AGN.  One marked 
example  is CGCG\,448-020, of which $L_{\rm IR}$ is dominated by 
an extremely compact, extranuclear starburst (Inami et al.~2010).  This 
might hint that a lower $R_{\rm midCO}$ could also be associated with 
some rare, extreme starbursts as a large-scale version of those Galactic 
extreme compact molecular cores where the CO line emission peaks at $J > 
10$ (Etxaluze et al.~2013; Habart et al.~2010).

Under the assumption that AGN/XDR results in CO line emissions mainly 
at $J > 10$, one can use Fig.~4 to estimate the fractional AGN contribution
to $L_{\rm IR}$ in an AGN where no significant Category-B shock is present:   
The measured $\log R_{\rm midCO} 
\approx \log R^{\rm SF}_{\rm midCO} + \log [1 - (L^{\rm AGN}_{\rm IR}/ 
L_{\rm IR})]$,
where $L^{\rm AGN}_{\rm IR}$ is the AGN component of $L_{\rm IR}$. 
For example, the observed $\log R_{\rm midCO}$ of $-4.45$ for Mrk\,231 implies 
that $L^{\rm AGN}_{\rm IR}/L_{\rm IR} \sim 52\%$ ($\pm 10\%$), in good 
agreement with $f_{\rm AGN} \approx 56\%$ from our mid-IR diagnostics.   
In the case of NGC\,1068, we have $L^{\rm AGN}_{\rm IR}/L_{\rm IR} \sim 
54\%$ ($\pm 10\%$), in good agreement with an independent estimate in 
Telesco \& Decher~(1988).

In summary, (1) we demonstrated that the SF-dominated LIRGs show a relatively
tight distribution in terms of $\log R_{\rm midCO}$, with a mean 
of $-4.13$ and a sample standard deviation of $0.10$; and (2) we showed,  
albeit with small number of cases, the possibility that (a) galaxies bearing 
powerful interstellar shocks not associated with current SF and (b) galaxies 
with a significant AGN contribution to their bolometric luminosity, have 
their $\log R_{\rm midCO}$ higher and lower than $-4.13$, respectively. 

While it is rather conclusive that the heating source of the mid-$J$ CO line 
emission in most LIRGs is current SF, more work needs to be done to clarify 
which SF-related mechanisms are directly responsible for the warm CO line emission.

\acknowledgments

This paper benefited from a number of useful comments made by an 
anonymous referee.
This work is based in part on observations made with \Herschel, 
a European Space Agency Cornerstone Mission with significant 
participation by NASA.  Support for this work was provided in 
part by NASA through an award issued by JPL/Caltech.  This 
research has made use of the NASA/IPAC Extragalactic Database (NED), 
which is operated by Jet Propulsion Laboratory, California 
Institute of Technology, under contract with NASA.


%

\newpage

\end{document}